\documentstyle[12pt]{article}

\newcommand{\tdm}[1]{\mbox{\boldmath$#1$}}
\newlength{\dinwidth}
\newlength{\dinmargin}
\def\lapproxeq{\lower .7ex\hbox{$\;\stackrel{\textstyle
<}{\sim}\;$}}
\def\gapproxeq{\lower .7ex\hbox{$\;\stackrel{\textstyle
>}{\sim}\;$}}

\begin{document}
\titlepage
\begin{flushright}
DTP/96/16  \\
March 1996 \\
\end{flushright}

\begin{center}
\vspace*{2cm}
{\Large \bf A model for $F_L$ and $R={F_L/F_T}$ at low $x$ and low $Q^2$}\\
\vspace*{1cm}
B. Bade\l{}ek,\\
Institute of Experimental Physics, Warsaw University, 00-681 Warsaw, Poland\\
and\\
Department of Physics, Uppsala University, 751 21 Uppsala, Sweden\\
\vspace*{0.5cm}
J.\ Kwieci\'{n}ski\footnote{On leave from Henryk
Niewodnicza\'{n}ski Institute of Nuclear Physics, 31-342
Cracow, Poland.},\\

A. Sta\'sto \footnote{ 
On leave 
from the Department of Physics, Jagellonian University, 30-059 Cracow, Poland.}\\     
Department of Physics, University of Durham, Durham, DH1 3LE,
England
\end{center}
\vspace*{5cm}
\begin{abstract}
A model for the longitudinal structure function $F_L$ 
at low $x$ and low $Q^2$ is presented, which includes the kinematical constraint 
$F_L \sim Q^4$ as $Q^2 \rightarrow 0$.  
It is based on the photon-gluon fusion mechanism 
suitably extrapolated to the region of low $Q^2$.   
The contribution of quarks having limited transverse 
momentum is treated phenomenologically assuming that it is described by
the soft pomeron exchange mechanism.
The ratio $R={F_L/(F_2-F_L)}$, with the  
$F_2$ appropriately extrapolated to the region of low $Q^2$, is also discussed.      
\end{abstract}

\newpage
\section{Introduction}
The longitudinal structure function $F_L(x,Q^2)$ which corresponds to the 
interaction of the longitudinally polarized virtual photon in the
 one--photon--exchange mechanism of inelastic lepton-nucleon scattering is a very 
interesting dynamical quantity since at least at low $x$ the dominant contribution
to it comes from gluons. The experimental 
determination of $F_L$ is in general difficult 
and requires a measurement of the $y$ dependence of the deep 
inelastic cross section for fixed $x$ and $Q^2$ where,  as usual, 
$y=(p_lp)/(qp), x=Q^2/(2pq)$ and $Q^2=-q^2$ with $p_l, p$ and $q$ denoting 
the four momentum of the incident lepton, the four momentum of the proton 
and four momentum transfer between the leptons respectively, see fig. 1.  
In the "naive" quark-parton model the  
structure function $F_L(x,Q^2)$ vanishes. To be precise,  
in the limit of large $Q^2$ 
 it is  proportional to $(<m^2>+<\kappa_T^2>)/Q^2$  
where $m$ is the quark mass and $\kappa_T$ is the transverse momentum of 
the quark 
which,  in the naive parton model, is by definition limited 
\cite{lp,pvlp,pvls}. 
This remains approximately valid in the leading logarithmic 
approximation where only 
collinear gluon emissions are considered. 
The limitation of the quark transverse momentum no longer holds in the next-to-leading 
logarithmic approximation when the point-like QCD interactions allow (with probability 
$\sim \alpha_s(Q^2)$)   
$<\kappa_T^2>$ to grow as $Q^2$ with increasing $Q^2$.
Thus $F_L(x,Q^2)$ acquires a leading twist contribution, 
proportional to $\alpha_s(Q^2)$ \cite{am,gr,rgr}.  At small $x$ 
it is driven by the gluons through the $g \rightarrow q \bar q$ transition 
and, in fact, $F_L$ can be used as a very useful quantity for a direct measurement 
of the gluon distribution in a nucleon \cite{cs}. 

\medskip
In the limit $Q^2 \rightarrow 0$ the structure function $F_L$ has to vanish 
as $Q^4$ (for fixed $2pq$).  This kinematical constraint eliminates potential 
kinematical singularities at $Q^2=0$  of the hadronic tensor $W^{\mu \nu}$ 
defining the virtual Compton scattering amplitude \cite{rmp}.  
It also reflects the simple physical fact that the total 
cross section $\sigma_L \sim F_L/Q^2$ describing the interaction of the 
longitudinally polarized virtual photons has to vanish in the real 
photoproduction limit.

\medskip   
The differential cross section for inelastic lepton-nucleon scattering is 
most frequently expressed in terms of the structure function $F_2(x,Q^2)$ and 
the ratio $R(x,Q^2)=F_L(x,Q^2)/F_T(x,Q^2)$ where $F_T(x,Q^2)=F_2(x,Q^2)-
F_L(x,Q^2)$ is the structure function associated with transversely polarized
virtual photons.   
Unfortunately
measurements of $R(x,Q^2)$ are difficult and thus scarce,
especially at low $x$ and/or low $Q^2$. 
For $x<$ 0.01 and/or $Q^2 <$0.5 GeV$^2$ the 
experimental information is particularly poor.
The measurements of $R$ are compiled in fig.2. Preliminary measurements
in the low $x$ region have been announced by the 
CCFR collaboration \cite{unkidurham} and by NMC \cite{nmcrnew}.
All the data indicate a small value of $R$ at moderate values of $x$ and an 
increase
of $R$ with decreasing $x$. The data collected in fig.2 come from experiments
carried out on different targets. This is justified since the
measured differences $R^A - R^d$ do not exhibit any
significant dependence on $x$ and they are consistent with zero, 
\cite{nmcr}.
All the data in fig.2 (except those of SLAC E140X)
 have been fitted to a common function (the curve
in fig.2), $R^{fit}$, which is constructed to extrapolate to theoretically 
reasonable
values outside the kinematical range of the data: namely at $x \rightarrow 0$,
$x \rightarrow 1$
and $Q^2\rightarrow \infty$. In particular the last constraint means that
$R^{fit}$ matches $R^{QCD}$, \cite{am}, at high $Q^2$.
The fit expression should not however be used for
$Q^2<$0.35GeV$^2$ which means that
no constraints are given on the behaviour of $R$ at low $Q^2$.
This is a serious deficiency in the procedure used to extract 
polarized and non--polarized structure functions from the experimental data,
especially in the calculation of radiative corrections. 
Thus the procedure commonly accepted in data analysis is to assume
$R(x,Q^2)=R(x,Q^2=$0.35 GeV$^2)$  with 100$\%$ uncertainty
for $Q^2<$0.35 GeV$^2$. The uncertainty of $R$ is then propagated into
a contribution (usually the dominant one) to the systematic error on the
extracted structure functions. The need for a more precise determination
of $R$ at low $Q^2$ is thus clearly seen.

\medskip
At large $Q^2$ and large 
$x$ ($x >0.1$) perturbative QCD describes reasonably well the available 
data on the ratio $R(x,Q^2)$ 
\cite{cteq}. It has however been noticed that in order to 
correctly accommodate experimental information from the moderately 
large $Q^2$ region, a significant "higher twist" contribution, which vanishes 
as $1/Q^2$ at large $Q^2$, 
has to be added as well \cite{unkidurham,miramontes,bodekunki}.    

\medskip
While the longitudinal structure function is (at least theoretically) fairly 
well understood at high $Q^2$ very little (if anything) is known 
about its possible extrapolation towards the region of low $Q^2$.  
This should be contrasted with the structure function $F_2(x,Q^2)$ 
for which several dynamically motivated extrapolations exist, cf. \cite{rmp}.  

\medskip
The purpose of this paper is  to provide an extrapolation 
of the structure function $F_L(x,Q^2)$ towards the region 
of moderate and low values of $Q^2$. This extrapolation will be based 
on the photon-gluon fusion mechanism, essential at low $x$ and
 suitably extended into the region of low $Q^2$.  
A similar description of both the structure functions $F_2$ and $F_L$ has been 
 considered in ref. \cite {nikolaev}.  
We shall explicitly introduce an additional 
 higher twist term which comes from the contribution of the 
quarks (antiquarks)  with limited transverse 
momentum.  We assume that this term is described by "soft" pomeron 
exchange. We treat this contribution 
phenomenologically and determine its normalisation  from the 
(non-perturbative) part of the structure function $F_2(x,Q^2)$.  
The model embodies the kinematical constraint $F_L \sim Q^4$ in the 
limit $Q^2 \rightarrow 0$ (for fixed $2pq$ or $\nu$). 

\medskip
The paper is organised as follows:
in the next section we discuss the photon--gluon fusion contribution to the 
structure function $F_L$ and its extension to the low $Q^2$ region.  
In section 3 we present a phenomenological model for the higher 
twist contribution to $F_L$ at small $x$ which is based on the 
soft pomeron exchange mechanism. Section 4 contains 
numerical predictions for $F_L$ and $R$, while section 5 
contains a brief summary of our results.

\medskip
\section{The photon - gluon fusion model for the structure
function $F_L$}
At small $x$ and at large $Q^2$ the structure functions can be
computed from the $k_T$ factorization theorem \cite{kt,ktr,ktbl} which 
corresponds to the (virtual) photon - gluon fusion mechanism
shown in fig.3.
The structure function $F_L$ obtained from the diagram of
fig.3 is given by \cite{akms1} : 
\begin{equation} 
F_L(x,Q^2)=2 {Q^4\over \pi^2} \sum _qe_q^2 I_q(x,Q^2)
\label{flsum} 
\end{equation}
where,
\begin{equation} 
I_q(x,Q^2)=\int {dk_T^2\over k_T^4} \int_0^1 d\beta \int d^2\kappa_T^{\prime} 
\alpha_s \beta^2 (1-\beta)^2 {1\over 2}({1\over D_{1q}} - {1\over D_{2q}})
^2f(y_q,k_T^2) 
\label{flint}
\end{equation} 
and 
$$ D_{1q}=\kappa_T^2 + \beta (1-\beta) Q^2 + m_q^2 $$
\begin{equation}
 D_{2q}=(\tdm{\kappa_T} - \tdm{k_T})^2 + \beta (1-\beta) Q^2 + m_q^2\;.
\label{d} 
\end{equation}
$\alpha_s$ is the QCD coupling and its
argument will be specified later.
The transverse momentum $\tdm{\kappa_T^{\prime}}$ and the variables
$y_{q}$ are defined as follows: 
\begin{equation}
\tdm{\kappa_T^{\prime}} = \tdm{\kappa_T} - (1-\beta)\tdm{k_T }
\label{kappap}
\end{equation}
\begin{equation}
 y_q=x\left(1 + {\kappa_T^{\prime 2} + m_q^2 \over \beta (1-\beta)Q^2} + 
{k_T^2\over Q^2}\right) 
\label{yq}
\end{equation}
where $\tdm{\kappa_T}$ is the transverse momentum of the quark(antiquark) in the quark
box (see fig.3). The variable $\beta$ is the corresponding Sudakov
parameter appearing in the quark momentum decomposition into the basic
lightlike four vectors,  $p'$ and $q'$,
\begin{eqnarray}
\kappa=x_qp'-\beta q'+\tdm{\kappa_T}  \nonumber \\
x_q=x\left(1+{m_q^2+\kappa_T^2 \over (1-\beta)Q^2}\right) \nonumber \\
p'=p-{M^2 x \over Q^2}q    \\ 
q'=q+xp  \nonumber
\label{decomp}
\end{eqnarray}
where $M$ denotes the nucleon mass.
The function $f(y,k^2)$ is the so -- called unintegrated gluon distribution.  
It is related to the (scale dependent) gluon distribution 
$g(y,\mu^2)$ by
\begin{equation} 
yg(y,\mu^2)=\int^{\mu^2}{dk_T^2\over k_T^2} f(y,k_T^2)\;.
\label{intg}
\end{equation} 

\medskip
The parameters $m_q$ are the masses of the quarks.  It should be noted that
the integrals $I_q$ defined by (\ref{flint})  are infrared finite
even if we set $m_q=0$.  
The non-zero values of the quark masses are however necessary if 
formula 
(\ref{flsum}) is extrapolated down to $Q^2=0$ respecting the 
kinematical constraint $F_L \sim Q^4$.  The non-zero quark masses then 
play the role of the infrared regulator.  
It should be noticed that the expressions (\ref{flsum},\ref{flint})
 define $F_L$ as an analytic function 
of $Q^2$ except for the cut $Q^2 <- 4m_q^2$.  
Equation (\ref{flint})
can be interpreted as representing the rescattering of the 
$q\bar q$ fluctuation of the virtual photon
through the exchange of two and more interacting gluons.
Both the diagonal 
and non--diagonal transitions in this rescattering are included.
We assume that equations (\ref{flsum},\ref{flint})   
embody (at least in the average sense and as far as the position 
of the singularities in the $Q^2$ plane are concerned) also the 
virtual vector meson contributions which couple to virtual photons.  
This will fix the magnitude of the corresponding quark mass parameters to be  
$m_q^2 \approx m_v^2/4$ where $m_v$ denotes the mass of the lightest vector
meson corresponding to the $q \bar q$ pair.  

\medskip
The integration limits in (\ref{flint}) are  constrained by the 
condition:
\begin{equation} 
y_q<1  
\label{y1}
\end{equation}
The condition $y_q>x$ is automatically satisfied (see eq.(\ref{yq}) )

\medskip
The unintegrated gluon distribution $f(x,k^2)$
should in principle be obtained from the solution of the 
Balitzkij--Fadin--Kuraev--Lipatov (BFKL)   
equation \cite{bfkl,lip}. In our calculations we shall use the following
approximation,
\begin{equation}
f(y,k_T^2)=\left. y\frac{\partial g^{AP}(y,Q^2)}{\partial lnQ^2}\right|_{Q^2=k_T^2}
\label{derivap}
\end{equation} 
where $g^{AP}(y,Q^2)$ satisfies the conventional (LO or NLO) Altarelli-Parisi equations.
In this approximation one neglects the higher order small $x$ resummation
effects in the
gluon anomalous dimension. The gluon anomalous dimension $\gamma_{gg}(\alpha_s,
\omega)$ is the moment of the splitting function $P_{gg}(z)$,
\begin{equation}
\gamma_{gg}(\alpha_s,\omega)=\int_0^1dz z^{\omega}P_{gg}(z,\alpha_s)\;.
\label{splitting}
\end{equation} 
The BFKL equation effectively resums the leading powers of ${\alpha_s/ 
\omega }$ i.e.  
\begin{equation}
\gamma_{gg}(\alpha_s,\omega)={\bar\alpha_s \over \omega}+
\sum_{n=4} D_n \left({\bar\alpha_s \over \omega}\right)^n 
\label{andim}
\end{equation}
$(\bar\alpha_s={3\alpha_s/ \pi})$,
where the coefficients $D_n$ can be calculated from the kernel of the BFKL
equation \cite{jaroszewicz}. It should be noted that the sum on the right--hand--side
of (\ref{andim}) only starts at $n=4$. As a result it affects the evolution
of gluon distribution (and of observable quantities) only at very small values
of $x$ ($x<10^{-4}$). In our approximation this contribution is neglected.

\medskip
Expression (\ref{flsum}) can be represented in the following
"$k_t$ factorized" form: 
\begin{equation}
F_L(x,Q^2)=\int_x^1 {dx^{\prime}\over x^{\prime}}\int{dk_T^2\over k_T^2}
 \hat F_L^{0}(x^{\prime},Q^2,k_T^2)f({x\over x^{\prime}},k_T^2)
\label{kt}
\end{equation}
where the function $\hat F_L^{0}(x^{\prime},Q^2,k_T^2)$ can be regarded
as the longitudinal structure function of the off-shell gluon
with virtuality $k_T^2$.
It corresponds to the quark box (and crossed-box) diagrams in the upper 
part of the diagram in fig.3
and is given by the following formula: 
\begin{equation}
\hat F_L^{0}(x^{\prime},Q^2,k_T^2)={Q^4\over \pi^2 k_T^2}\sum_{q}e_q^2
 \int_0^1 d\beta \int d^2\kappa_T^{\prime}x^{\prime}\delta 
\left(x^{\prime} -  \left(1 + {\kappa_T^{\prime 2} + m_q^2 \over \beta (1-\beta)Q^2} + 
{k_T^2\over Q^2}\right)^{-1}\right)\times
\nonumber
\end{equation}
\begin{equation}
\times\alpha_s \beta^2 (1-\beta)^2 ({1\over D_{1q}} - {1\over D_{2q}})
^2
\label{flhat}
\end{equation}
where $D_{1q}, D_{2q}$ are defined by equation (\ref{d}).
The leading twist part of the $k_t$ factorization formula can be rewritten 
in a collinear factorized form:
\begin{equation}
F_L(x,Q^2)=\int_x^1 {dx' \over x'}C_L(x',Q^2,\alpha_s(Q^2)){x'\over x}
g({x' \over x},Q^2)
\label{collinear}
\end{equation}
where the leading powers of $\alpha_s ln(x')$, \, ($\alpha_s ln(x/x')$)
have been resummed in $C_L(x',Q^2,\bar\alpha_s(Q^2))$, \,$(x/x'g(x/x',Q^2))$.
Thus for instance the moment function 
\begin{equation}
\bar C_L(\omega,Q^2,\alpha_s(Q^2))\equiv\int_0^1{dz \over z} z^{\omega}
C_L(z,Q^2,\alpha_s(Q^2))
\end{equation}
is related to the moment function $\tilde F_L^{0}(\omega,\gamma)$, by
\begin{equation}
\bar C_L(\omega,Q^2,\alpha_s(Q^2))=\gamma_{gg}(\alpha_s,\omega)
\tilde F_L^{0}(\omega,\gamma=\gamma_{gg}(\alpha_s,\omega))\;,
\end{equation}
where
\begin{eqnarray}
\bar F_L^{0}(\omega,Q^2)&=&
{1 \over 2\pi i}\int_{1/2-i\infty}^{1/2+i\infty}d\gamma
\tilde{F_L^{0}}(\omega,\gamma)
\left({Q^2 \over k_T^2}\right)^{\gamma}   \\ \nonumber 
\bar F_L^{0}(\omega,Q^2)&=&\int_0^1{dx \over x}x^{\omega} \hat{F_L^{0}}(x,Q^2)\;.
\end{eqnarray}
Our approximation corresponds to setting $\gamma_{gg}=\bar\alpha_s/\omega$.
It correctly generates the first three corrections to $C_L$ which are presumably most
important at moderately small values of $x$.

\medskip
In the  "on-shell" approximation  one sets $k_T^2=0$ in the argument of 
$\hat{F_L^0}(x',Q^2,k_T^2)$ and restricts the integration over $k_T^2$ to the region
$k_T^2\ll \kappa'^2_T \sim Q^2$. After taking into account (\ref{intg}) the structure function
$F_L$ can be expressed in terms of the conventional gluon distribution $yg(y,Q^2)$,
\begin{equation}
I_q(x,Q^2)= \pi \int_0^1 d\beta  \int d\kappa_T^{\prime 2} 
\alpha_s(Q^2) \beta^2 (1-\beta)^2 {\kappa_T^{\prime 2}\over D_q^4}
y_qg(y_q,Q^2) 
\label{iq}
\end{equation}
where now
\begin{equation}
y_q=x\left(1 + {\kappa_T^{\prime 2} + m_q^2 \over \beta (1-\beta)Q^2}\right) 
\label{y}
\end{equation}
and
\begin{equation}
D_q= \kappa_T^{ \prime 2} + \beta (1-\beta) Q^2 + m_q^2  \;.
\label{dq}
\end{equation}
\par
In order to extrapolate the formula (\ref{flsum}) to the low $Q^2$ region
(with $I_q$ given by (\ref{flint}) or (\ref{iq})) 
we have to freeze the evolution of $g(y,Q^2)$ and freeze the argument 
of $\alpha_s(Q^2)$.  To this end we  
  substitute $Q^2+4m_q^2$ instead of $Q^2$ as the argument of
$\alpha_s$ and of $yg(y,Q^2)$.    
This substitution may be justified by analyticity arguments i.e. 
we want $I_q$ to have "threshold" singularities for $Q^2< -4m_q^2$.  
Moreover, for heavy quarks, it is the heavy mass squared and not $Q^2$ 
which should control  the scale of $\alpha_s$ and of the parton 
distributions for moderately large values of $Q^2$.
It should be noted that after those modifications the 
structure function $F_L$ can be continued down to $Q^2=0$,
respecting the kinematical constraint
( $F_L \sim Q^4$ in the limit $Q^2 \rightarrow 0$ ). 

\medskip 
It follows from (\ref{iq})  that the structure function $F_L$ in the on-shell
approximation can be represented by 
\begin{equation}
F_L=2\sum_q e_q^2 (J_q^{(1)}-2{m_q^2 \over Q^2} J_q^{(2)})
\label{efel}
\end{equation}
where
\begin{equation} 
J_q^{(1)}={\alpha_s \over \pi} \int_{\bar x_q}^1{dy\over y}
\left({x\over y}\right)^2
\left(1-{x\over y}\right)\sqrt{1-{4m_q^2 x\over Q^2(y-x)}} 
yg(y,Q^2)
\label{iq3}
\end{equation}
\begin{equation}
J_q^{(2)}={\alpha_s \over \pi} \int_{\bar x_q}^1
{dy\over y}\left({x\over y}\right)^3
ln \left({1+\sqrt{1-{4m_q^2 x\over Q^2(y-x)}}\over 
1-\sqrt{1-{4m_q^2 x\over Q^2(y-x)}}}\right)yg(y,Q^2) 
\label{iq4}
\end{equation}
and
\begin{equation}
\bar x_q=x(1+{4m_q^2 \over Q^2})\;.
\label{barxq}
\end{equation}
It should be noted that in the large $Q^2$ limit, the first term $J_q^{(1)}$ 
becomes equal to the standard QCD formula describing the leading order
contribution to $F_L$ originating from the photon-gluon fusion \cite{am,rgr},
\begin{equation}
F_L(x,Q^2)=2{\alpha_s \over \pi}\sum_q e_q^2 \int_x^1({x \over y})^2(1-{x \over y})
yg(y,Q^2)\;.
\label{altmar}
\end{equation}
\noindent
In the large $Q^2$ region formula (\ref{efel}) with non-zero quark masses
generates also power corrections enhanced by logarithmic factor
present in the term $J_q^{(2)}$.

\medskip
When analysing the $k_T$ factorization formula it is convenient to use
the "on-shell" approximation in the region $k^2<k_0^2$, where $k_0^2$
is a parameter of the order 1 GeV$^2$ and to leave the remaining part
($k^2>k_0^2$) unchanged. The first term is then given by (\ref{efel}), 
(\ref{iq3}), (\ref{iq4})
where gluon distribution is evaluated at the scale $k_0^2$ instead of $Q^2$
\cite{ktbl}. This gives the following representation for the structure
function $F_L$:
$$F_L(x,Q^2)=F_{L0}(x,Q^2;k_{T0}^2)+$$ 
\begin{equation}
2{Q^4 \over \pi^2}\sum_q e_q^2 \int 
{dk_T^2\over k_T^4} \Theta(k_T^2-k_{T0}^2)
\int_0^1 d\beta \int d^2\kappa_T^{\prime} 
\alpha_s \beta^2 (1-\beta)^2 {1\over 2}({1\over D_{1q}} - {1\over D_{2q}})
^2f(y_q,k_T^2)
\label{flmod}
\end{equation}
where $F_{L0}(x,Q^2;k_{T0}^2)$ is given by equations
(\ref{efel}) 
-- (\ref{iq4}) with $yg(y,k_{0T}^2)$ instead of $yg(y,Q^2)$
in the integrals (\ref{iq3}) and (\ref{iq4}). The argument of $\alpha_s$
in the formulae (\ref{iq3}), (\ref{iq4}) defining 
$F_L(x,Q^2)=F_{L0}(x,Q^2;k_{T0}^2)$ and in the second term of (\ref{flmod})
will be set equal to $Q^2+4m_q^2$.
\medskip
\section{A model for the higher twist contribution to $F_L$}
The phenomenological description of the experimental 
data for the ratio $R(x,Q^2)$ in the region of moderately 
large values of $Q^2$ and large 
values of $x$ ($x>0.1$) requires a significant higher twist contribution, 
i.e. the term which vanishes as $1/Q^2$ for $Q^2 \rightarrow \infty$ 
\cite{unkidurham,miramontes,bodekunki}.

\medskip
In this section we present the phenomenological model for $F_L$ which 
takes into account the higher twist effect. The idea is to treat the
contributions from the regions of low and high values of 
quark transverse momenta in  different ways.
That is we divide the
integration over $\kappa'^2_T$ into two parts: the region
 $0<\kappa'^2_T<\kappa'^2_{0T}$, and $\kappa'^2_T>\kappa'^2_{0T}$,
  where $\kappa'^2_{0T}$ is
an arbitrary  cut-off, chosen to be of the order of 1 GeV$^2$.

\medskip 
   We leave unchanged the contribution coming from the high
$\kappa_T^{\prime 2}$ region whereas in the low $\kappa_T^{\prime 2}$
region, which is presumably dominated by the soft physics, we use
the "on-shell" approximation and  make
the substitution: 
\begin{equation}
\alpha_s zg(z,Q^2) \rightarrow A\;.
\label{subst}
\end{equation}
This gives the following representation of the higher twist
contribution to $F_L$: 
\begin{equation}
F_L^{HT}=2A\sum_q e_q^2 {Q^4\over \pi}
\int_0^1 d\beta \beta^2 (1-\beta)^2 \int_0^{\kappa^{\prime 2}_{0T}}
 d\kappa_T^{\prime 2}{\kappa_T^{\prime 2}\over D_q^4}
\label{flht}
\end{equation} 
where $D_q$ has been defined in (\ref{dq}).
The constant $A$ is not a free parameter. We estimate it from $F_2$
assuming that the non-perturbative contribution to $F_2$ in the
scaling region also comes from the region of low values of $\kappa_T'^2$
and is controlled by the same parameter.
The term $F_L^{HT}$ can be interpreted as representing the contribution 
of soft pomeron exchange with intercept equal to 1. 

\medskip 
It should be noted that $F_L^{HT}$ given by equation (\ref{flht})
 will vanish as $1/Q^2$ in the high $Q^2$
limit (modulo logarithmically varying factors).  We call it
therefore a "higher twist contribution".  It should be
noticed that this term will also respect the kinematical
constraint $F_L \sim Q^4$ in the limit $Q^2 \rightarrow 0$. 

\medskip
The corresponding formula describing the photon - gluon
contribution to the structure function $F_T$ is
\begin{eqnarray}
F_T(x,Q^2)&=&2\sum_qe_q^2 {Q^2\over 4 \pi} \alpha_s
\int_0^1 d\beta \int d\kappa_T^{\prime 2} {x\over x^{\prime}}
g({x\over x^{\prime}},Q^2) \times \nonumber \\
&& \times\left[{\beta^2 + (1 -
\beta)^2 \over 2} \left ({1\over D_q^2}- {2 \kappa_T^{\prime 2}
\over D_q^3} +{2\kappa_T^{2}\kappa_T^{\prime 2} \over D_q^4}\right) 
+{m_q^2 \kappa_T^{\prime 2}\over D_q^4} \right]\;.
\label{ft}
\end{eqnarray}
We integrate this expression over the low $\kappa_T^{\prime 2}$ region
($\kappa_T^{\prime 2}<\kappa_{0T}^{\prime 2}$)
after substituting $\alpha_s z g \rightarrow A$ and retain 
its limiting value as $Q^2 \rightarrow \infty$.  We identify this
contribution 
to $F_T$ with the "background" term $F_2^{Bg}$ of ref.\cite{akms}
i.e.: 
\begin{equation} 
F_2^{Bg}=A\times{\sum_qe_q^2\over \pi} \int _0^{\infty} dt
 \int _0^{\kappa_{0T}^{\prime 2}}d\kappa_T^{\prime 2} 
\left[{1 \over 2} \left ({1\over D_q^2}- {2 \kappa_T^{\prime 2}
\over D_q^3} +{2\kappa_T^{2}\kappa_T^{\prime 2} \over D_q^4}\right) 
+{m_q^2 \kappa_T^{\prime 2}\over D_q^4} \right]     
\label{ftht}
\end{equation}
where now 
\begin{equation}
D_q=m_q^2 + \kappa_T^{\prime 2} + t \;.
\label{dqft}
\end{equation}
A possible weak $x$ dependence of $F_2^{Bg}$ is neglected and we set 
$F_2^{Bg}=0.4$.

\medskip
The complete structure function $F_L$ is represented as 
$F_L=F_L^{HT}+F_L^{LT}$
where $F_L^{LT}$ is calculated from equations (\ref{efel})
-- (\ref{iq4}), or from equation (\ref{flmod}) with the additional constraint
$\kappa'^2_{T}>\kappa'^2_{0T}$ in order to avoid double counting.

\medskip

\section{Numerical results}
In this section we present the numerical analysis of the structure
function $F_L$. We shall also perform the analysis of the ratio  
$R(x,Q^2)$ and compare it with the available data.

\medskip
In our calculations we have included the contributions of the $u,d,s,c$ quarks
with masses $0.35, 0.35, 0.5, 1.5 GeV$ respectively.
The parameter $\kappa'^2_{0T}$ was set to be equal 0.8 GeV$^2$.
We have varied this parameter in the interval 0.8--1.5 GeV$^2$ and found
that the results for $R$ are not very sensitive to these changes,
i.e. the ratio $R$ does not change by more than 10\%.

\medskip 
We have used the GRV \cite{grv} and MRS(A) \cite{mrsa} parton distributions 
both in the "on-shell" and "off-shell" approximation.
In the case of the GRV parametrisation
 LO gluons and QCD coupling constant were used.
In our calculations we froze
the evolution in $Q^2$ in the argument of $\alpha_s(Q^2)$  and also
in the function $yg(y,Q^2)$ as explained before. 
Besides the dominant (at small $x$) photon-gluon fusion
$\gamma g \rightarrow q\bar{q}$ contribution, 
we have also included in $F_L$ a 
contribution from quarks  taking into account threshold effects 
in the corresponding convolution integral i.e.
\begin{equation}
\Delta F_L(x,Q^2)=\sum_i e_i^2 \frac{\alpha_s(Q^2+4m_q^2)}{\pi}
\frac{4}{3}\int_{\bar x_q}^1\frac{dy}{y}
(\frac{x}{y})^2 [q_i(y,Q^2+4m_q^2)+\bar q_i(y,Q^2+4m_q^2)]\;.
\end{equation}
where $\bar x_q$ is given by (\ref{barxq}).

\medskip
In figures 4 and 5 we show our results for the longitudinal structure
function calculated without an additional higher twist contribution.
$F_L$ is plotted as the function of
$Q^2$ for different values of $x$, 
and as a function of $x$ for different values of $Q^2$.
The theoretical curves reflect the different behaviour of the 
MRS(A) and GRV gluon distributions.  The difference between the on--shell and off--shell
approaches is in general small, particularly in the small $Q^2$ region.
It can be seen from fig.5, that the magnitude of $F_L$ strongly decreases 
as $Q^2 \rightarrow 0$.

\medskip
In figures 6 and 7 we show results which include the additional higher twist
contribution. It can be seen from these plots that the resulting magnitude
of the longitudinal structure function is now bigger in the low $Q^2$ region
than in the previous case. It should also be noted that the structure 
function $F_L$ flattens as a function of $x$ in the low $x$ and
low $Q^2$ region. This is a direct consequence of the "soft pomeron"
parametrization of the higher twist term.

\medskip
We have also calculated the ratio $R(x,Q^2)=F_L(x,Q^2)/
(F_2(x,Q^2)-F_L(x,Q^2))$. The structure function $F_2$ was parametrized
following the model of ref. \cite{bbjk}.
In this model the structure function $F_2$ is represented by
\begin{equation}
F_2(x,Q^2)={Q^2 \over 4\pi}\sum_v{M_v^4\sigma_v(s) \over \gamma_v^2
(Q^2+M_v^2)^2}+{Q^2 \over Q^2+Q_0^2}F_2^{AS}(\bar x,Q^2+Q_0^2)\;.
\label{f2vdm}
\end{equation}
The first term describes the VMD contribution and the sum extends
over the low mass vector mesons. The parameter $\gamma_v$ can be estimated
from the leptonic widths of the vector mesons, $\sigma_v(s)$ is the total
vector meson - proton cross section, and $s=2pq-Q^2+M^2$ where $M$ is
the nucleon mass. The structure function $F_2^{AS}(x,Q^2)$ is obtained
from the QCD improved parton model. The variable $\bar x$ is defined
as 
\begin{equation}
\bar x=x(1+{Q_0^2 \over Q^2})\;.
\end{equation}
In the case of the GRV parametrization only the light quarks contribution
was included in $F_2^{AS}$ in the formula (\ref{f2vdm}).
The charm quark contribution to $F_2$ was calculated from the on-shell 
photon-gluon fusion.

\medskip
In figures 8 and 9 we show our results for the ratio $R(x,Q^2)$.
We see that this quantity is sensitive to the different parton
distributions which were used in calculating both $F_L$ and $F_2$.
In particular the structure function $F_2$ obtained from the formula
(\ref{f2vdm})
with $F_2^{AS}$ calculated from the MRS(A) parametrization turns
out to be steeper function of $x$ in the low $Q^2$ region than that
when $F_2^{AS}$ is calculated from the GRV partons.
As a result the ratio
$R$ decreases with $x$ as $x \rightarrow 0$. This effect is weaker
if $F_L$ does not contain the higher twist modification.
The spread of the theoretical predictions can be used as an estimate
of the corresponding errors caused by the uncertainty of $R$ in the
low $Q^2$, low $x$ region.
For the intermediate values of $Q^2$ this model coincides with
SLAC parametrization, whereas for lower values of $Q^2$, outside
the range of the validity of this parametrization, it quickly vanishes
as $Q^2 \rightarrow 0$. 

\medskip
In fig. 10 we show extrapolation of our model calculations for
$R$ up to $x=0.1$ and confront
this extrapolation with the available data. Although this value
of $x$ is already too large for our model to be applicable one can see
that it predicts reasonably well the magnitude and the $Q^2$ dependence
of $R$. The agreement with experimental results
improves if we renormalise the phenomenological
parameter $A$ by a factor of 1.2 -- 1.5 .
\section{Summary and conclusions}

In this paper we have presented possible parametrization of the longitudinal 
structure function $F_L$ and of the ratio $R(x,Q^2)$ in the region of 
low $Q^2$ and low $x$.  We based this parametrization on the photon-gluon 
fusion mechanism suitably extrapolated into the region of low $Q^2$. 
The extrapolation respected the kinematical constraint $F_L \sim Q^4$ 
in the limit $Q^2 \rightarrow 0$.   
We have also included a separate contribution from the higher twist term which 
corresponds to soft pomeron exchange.  Its coupling to external virtual 
photons was assumed to be proportional to that of the (on-shell) gluons 
but with limited transverse momentum of the quarks (antiquarks) within the 
quark box (and crossed-box) diagram, eq. (\ref{flht}).
We have also calculated the ratio 
$R(x,Q^2)=F_L(x,Q^2)/F_T(x,Q^2)$ utilising the parametrization of the structure 
function $F_2(x,Q^2)$ from ref. \cite{bbjk}.  Our results show that 
the structure function $F_L$ and the ratio $R$ become 
negligibly small in the region 
of low $Q^2$ ($Q^2 < $0.1 GeV$^2$ or so).  
The ratio $R(x,Q^2)$ turns out to be 
essentially independent of $x$ in the low $x$, low $Q^2$ region.  
We have also noticed that in this region the magnitude of both $F_L$ and 
of the ratio $R$ is relatively insensitive to the variations of the input 
parton distributions.  
We believe that our results will also be useful in the radiative corrections 
procedure as well as in extracting the magnitude of the 
spin dependent structure functions from the experimental data.
The codes for calculating $R(x,Q^2)$ are available upon request from
a.m.stasto@durham.ac.uk.
\section{Acknowledgments}

We thank Alan Martin for critically reading the 
manuscript and useful comments and Un-Ki Yang for discussions.
JK and AS thank the Physics Department and Grey College of the 
University 
of Durham as well as the Theoretical Physics Department at DESY  
for their warm hospitality.  This research has been supported 
in part by  the Polish KBN grant number 2 P302 062 04, by the Volkswagen Stiftung 
and by the 
EU contract number CHRX-CT92-0004/CT93-357.

\medskip\medskip\medskip
\vskip1cm
\noindent
{\Large {\bf Figure captions}}
\medskip\medskip
\begin{enumerate}
\item
Kinematics of inelastic charged lepton--proton scattering in the 
one--photon--exchange approximation and its relation through the
optical theorem to Compton scattering for the virtual photon;
$p_l$, $p$ and $q$ denote the four momenta of the incident lepton,
proton and virtual photon respectively.
\item
Compilation of measurements of $R$; data come from 
SLAC E140X \cite{e140x}, SLAC E140 and the global analysis of
earlier SLAC experiments \cite{whitlow}, CDHSW \cite{cdhsw}, BCDMS
\cite{bcdms} and EMC \cite{emc} experiments. Errors show the
statistical and systematic uncertainties added in quadrature but
in case of SLAC E140X and the global analysis of the SLAC data they 
do not include the systematic error due to radiative corrections
(equal to approximately 0.012 and 0.025 respectively).
The curve is a fit to all of the data (excluding E140X), \cite{whitlow,slacr}.
The figure is taken from \cite{unkidurham}.
\item
The diagramatic representation of the photon--gluon fusion mechanism
and of the $k_T$ factorization formula (\ref{kt}).
\item
Longitudinal structure function $F_L$ calculated as a function of $x$
for different values of $Q^2$ assuming the MRS(A) and GRV gluon 
distributions for both the on- and off-shell approaches.
\item
As fig.4 but $F_L$ is calculated as a function of $Q^2$ for different
values of $x$.
\item
As fig.4 but results include a contribution from higher twist.
\item
As fig.5 but results include a contribution from higher twist.
\item
$R(x,Q^2)$ plotted as a function of $x$ for different values of 
$Q^2$ and containing a contribution from higher twist.
\item
As fig.8 but $R$ is plotted as a function of $Q^2$ for different
values of $x$.
\item
Comparison of the model calculation for $R$, extrapolated to $x$=0.1,
with the data. The higher twist is taken into account in the calculations.
See fig. 2 for a description of the data.
\end{enumerate}

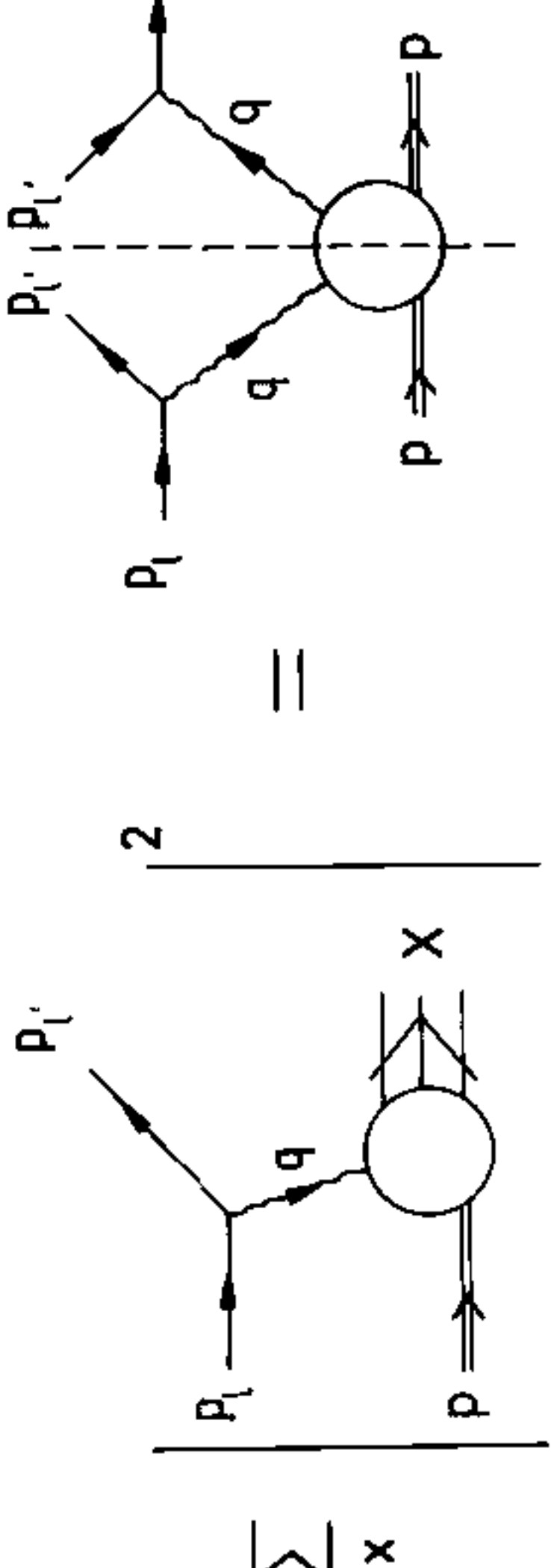
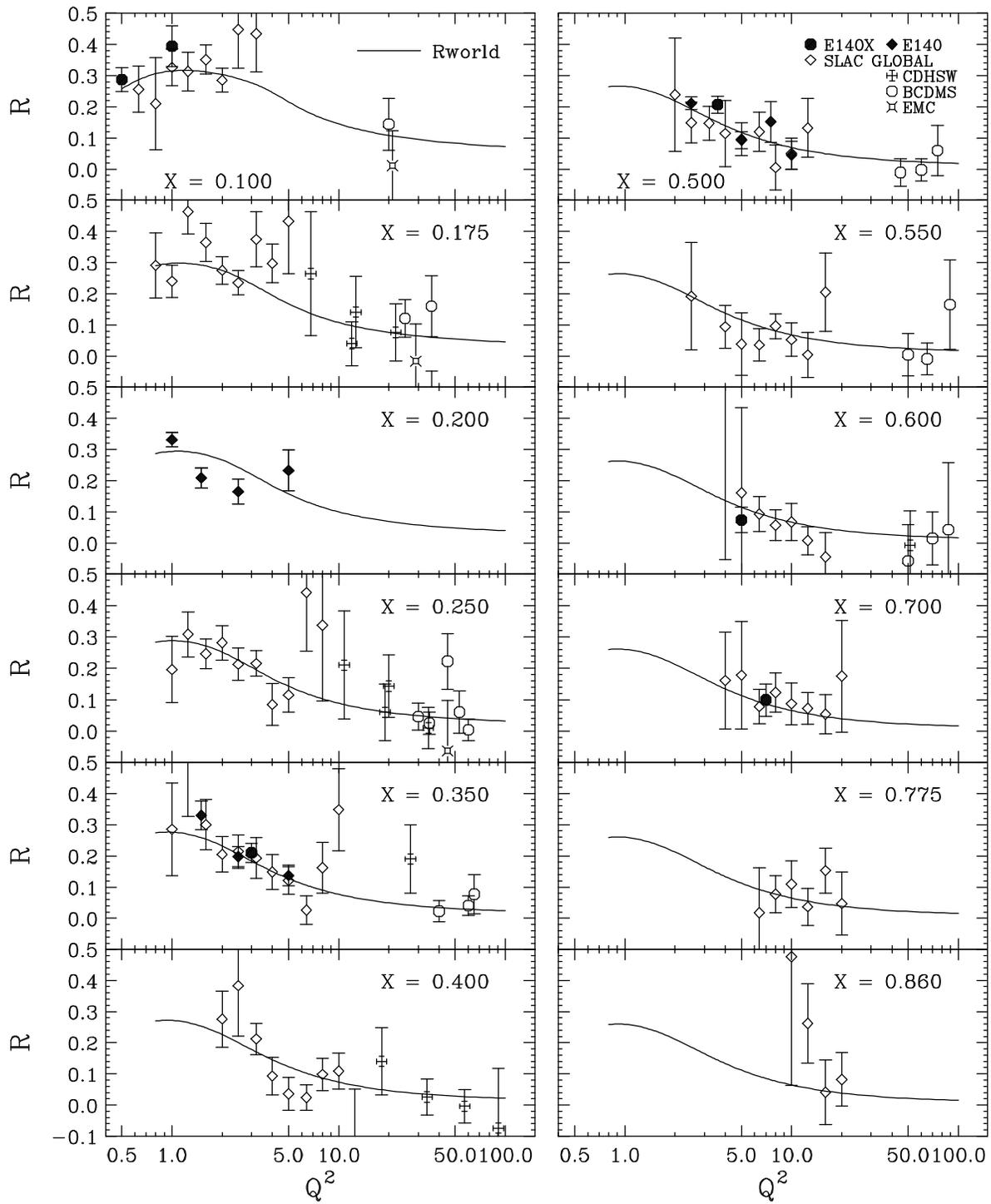
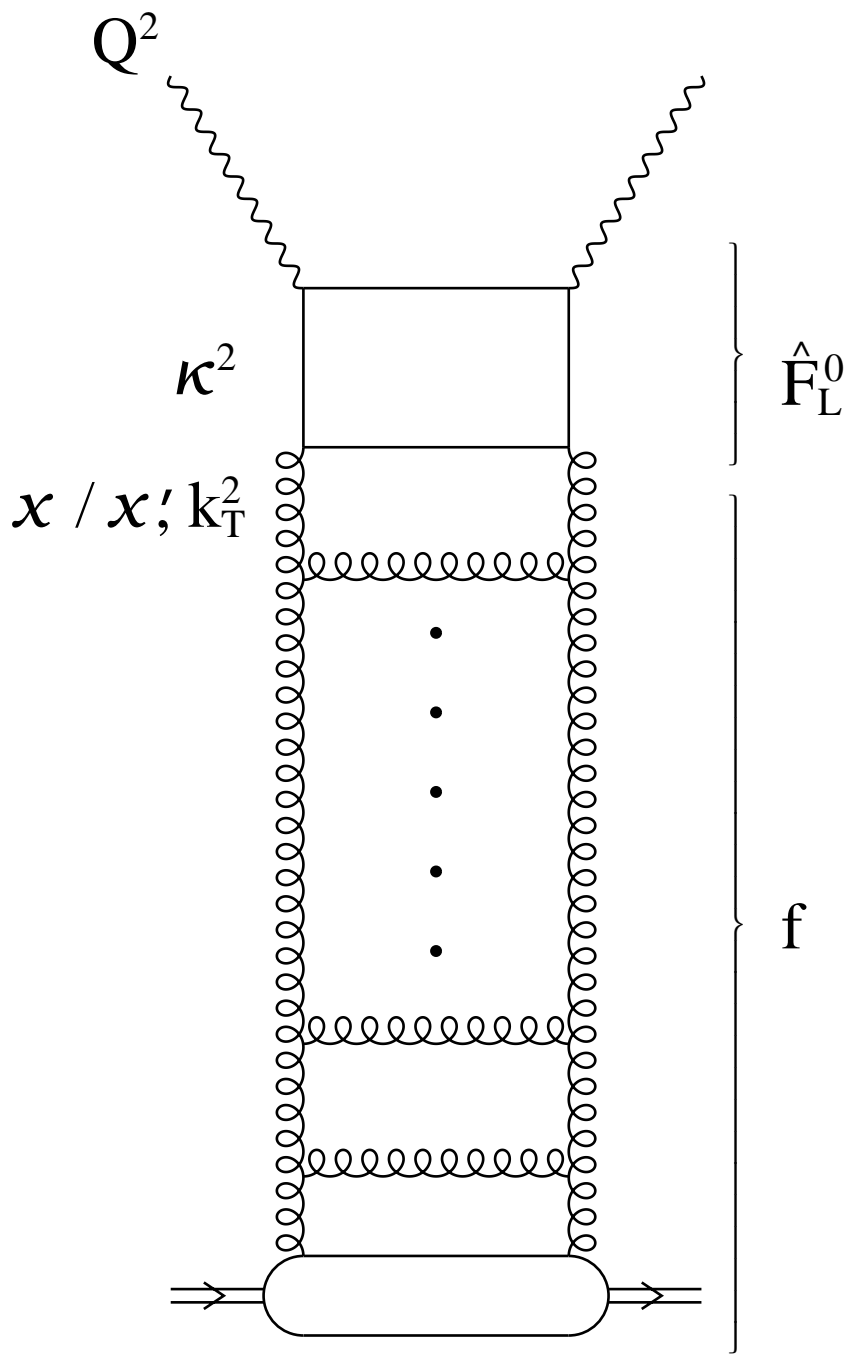
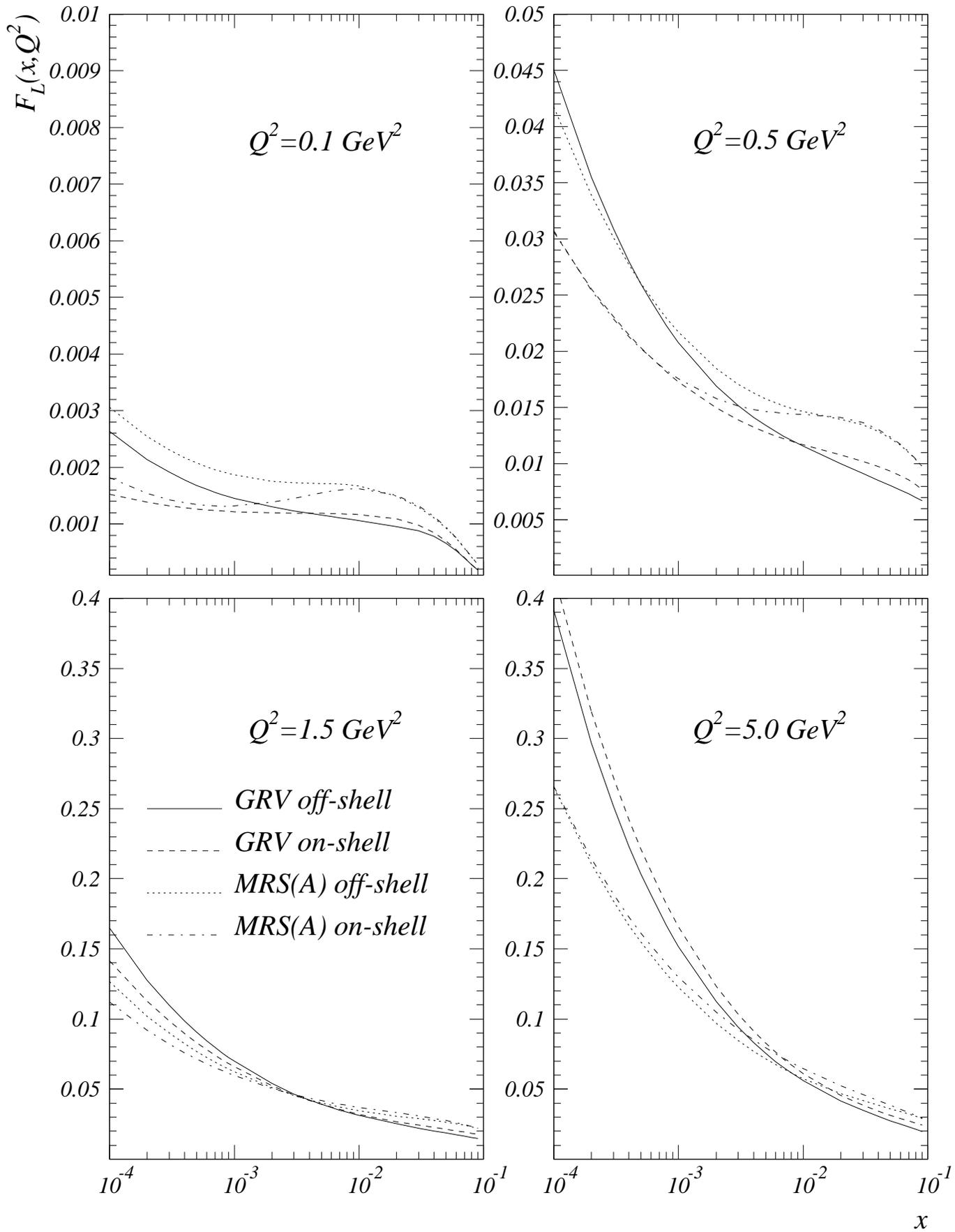
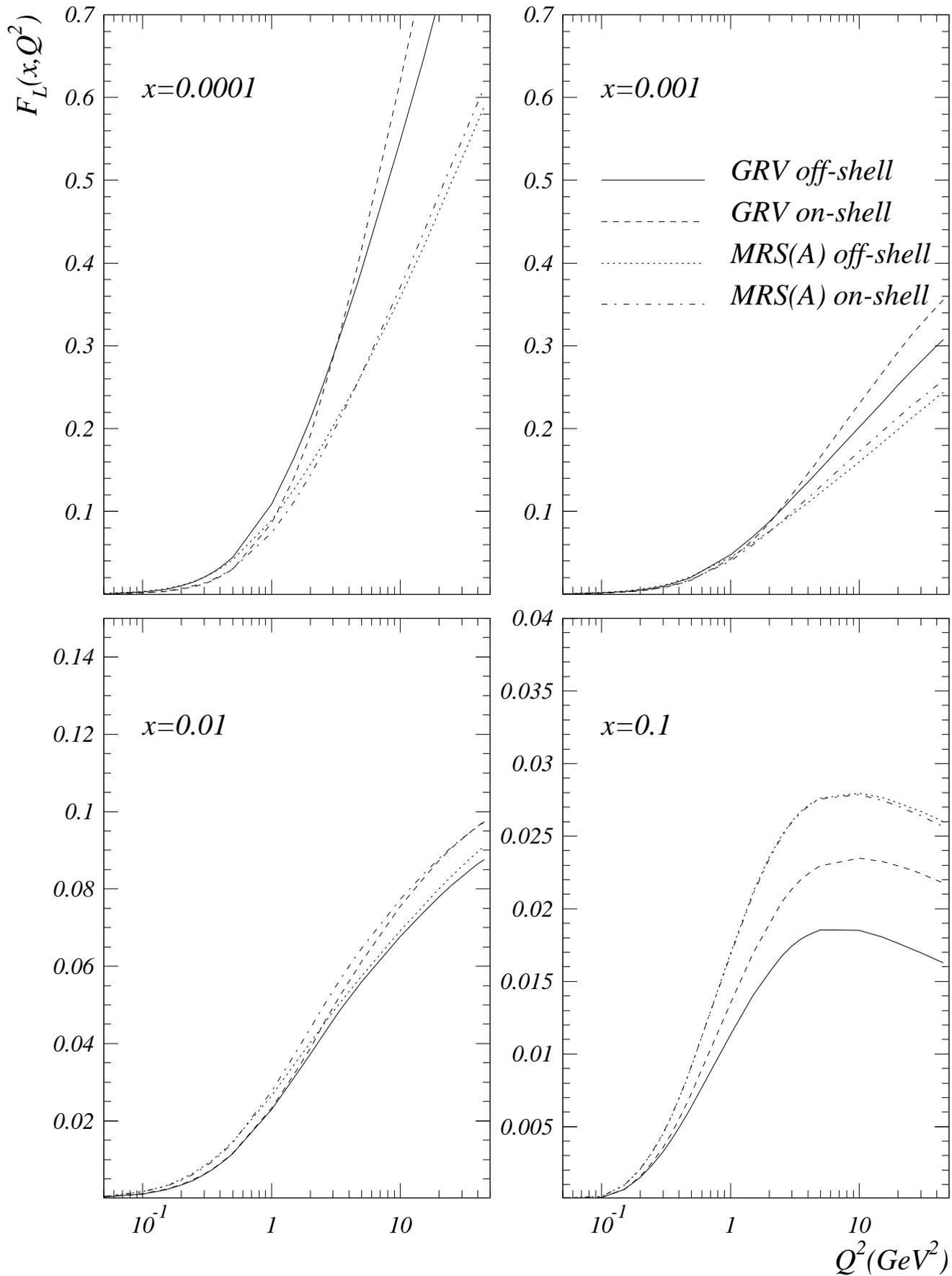
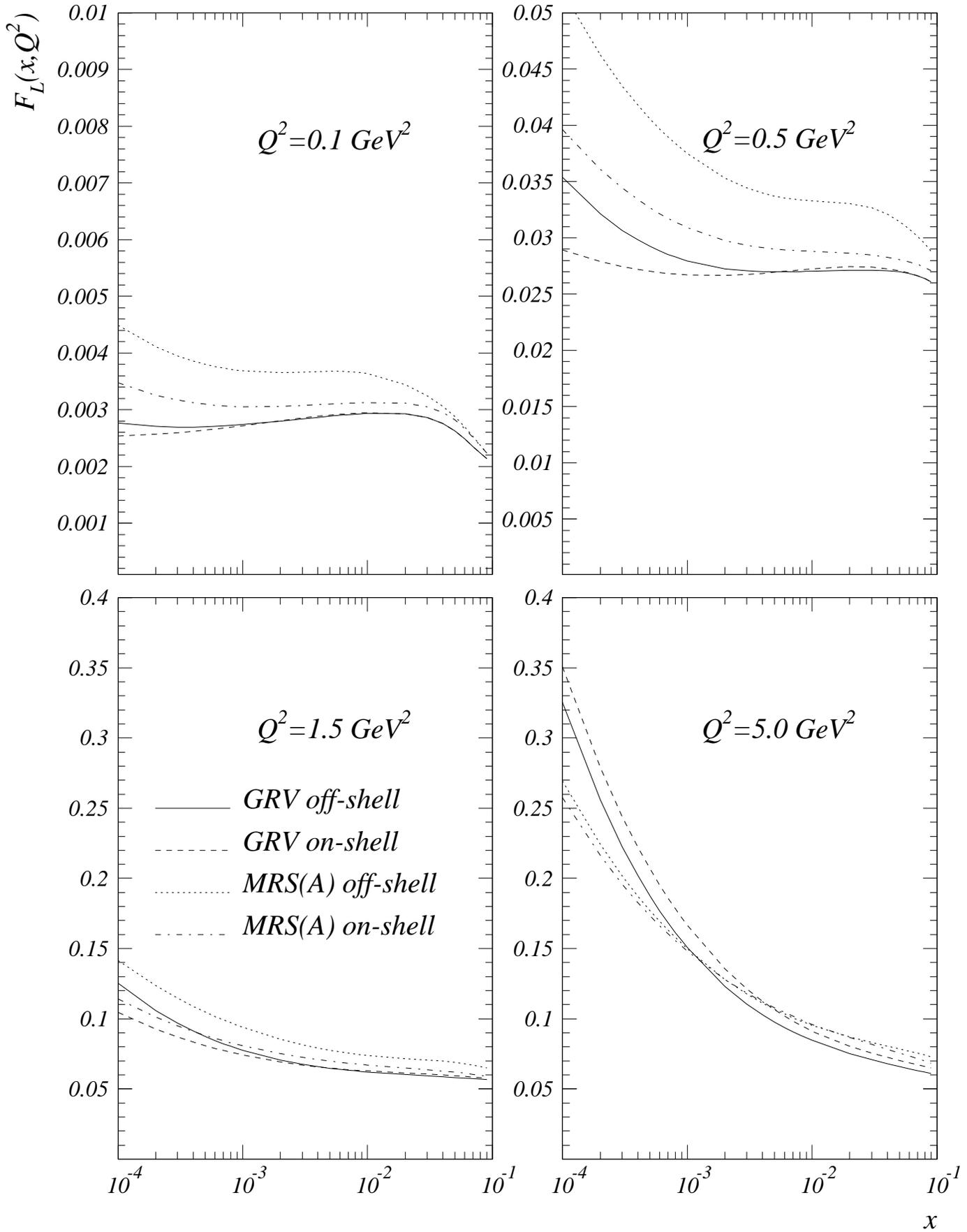
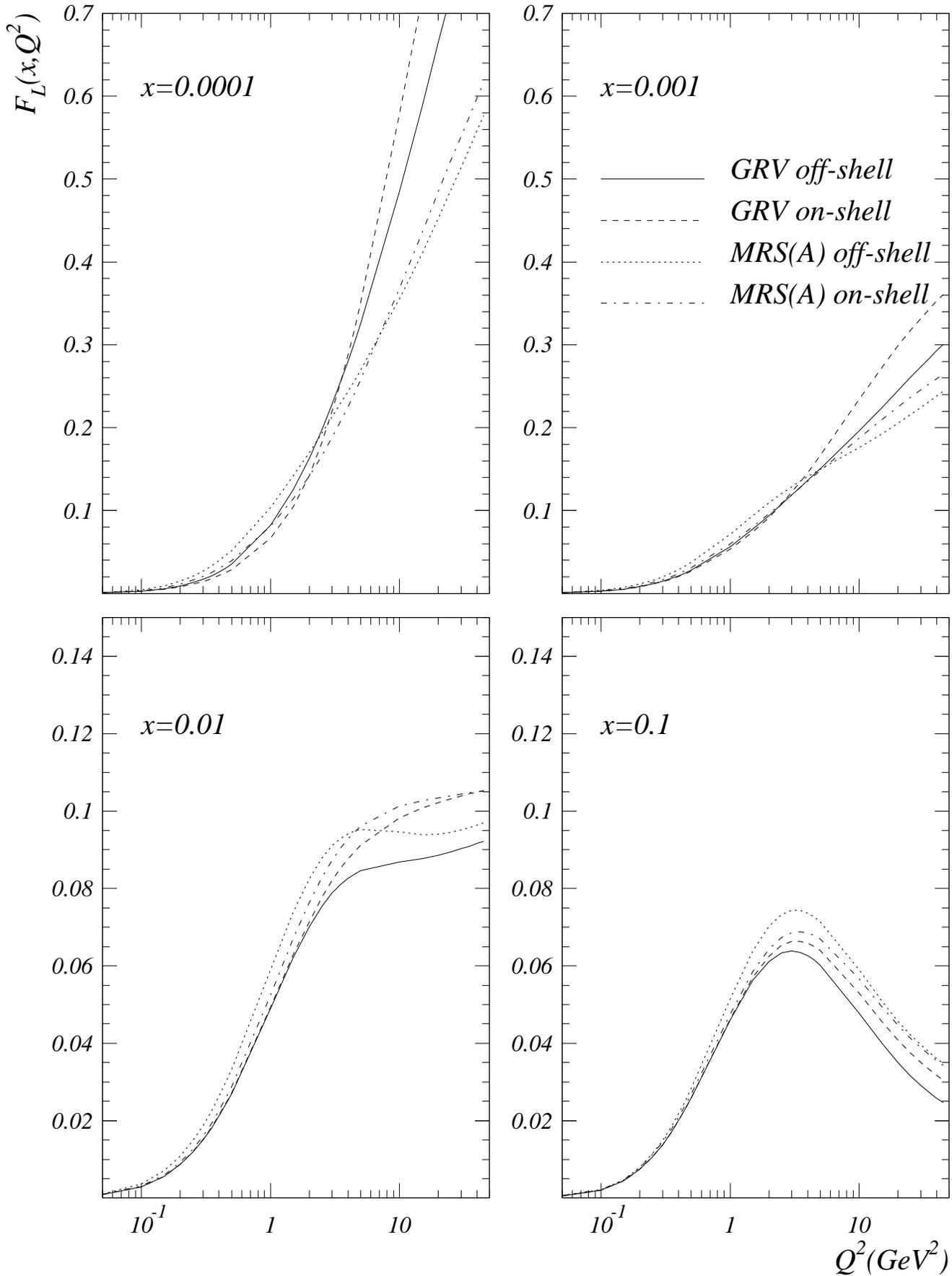
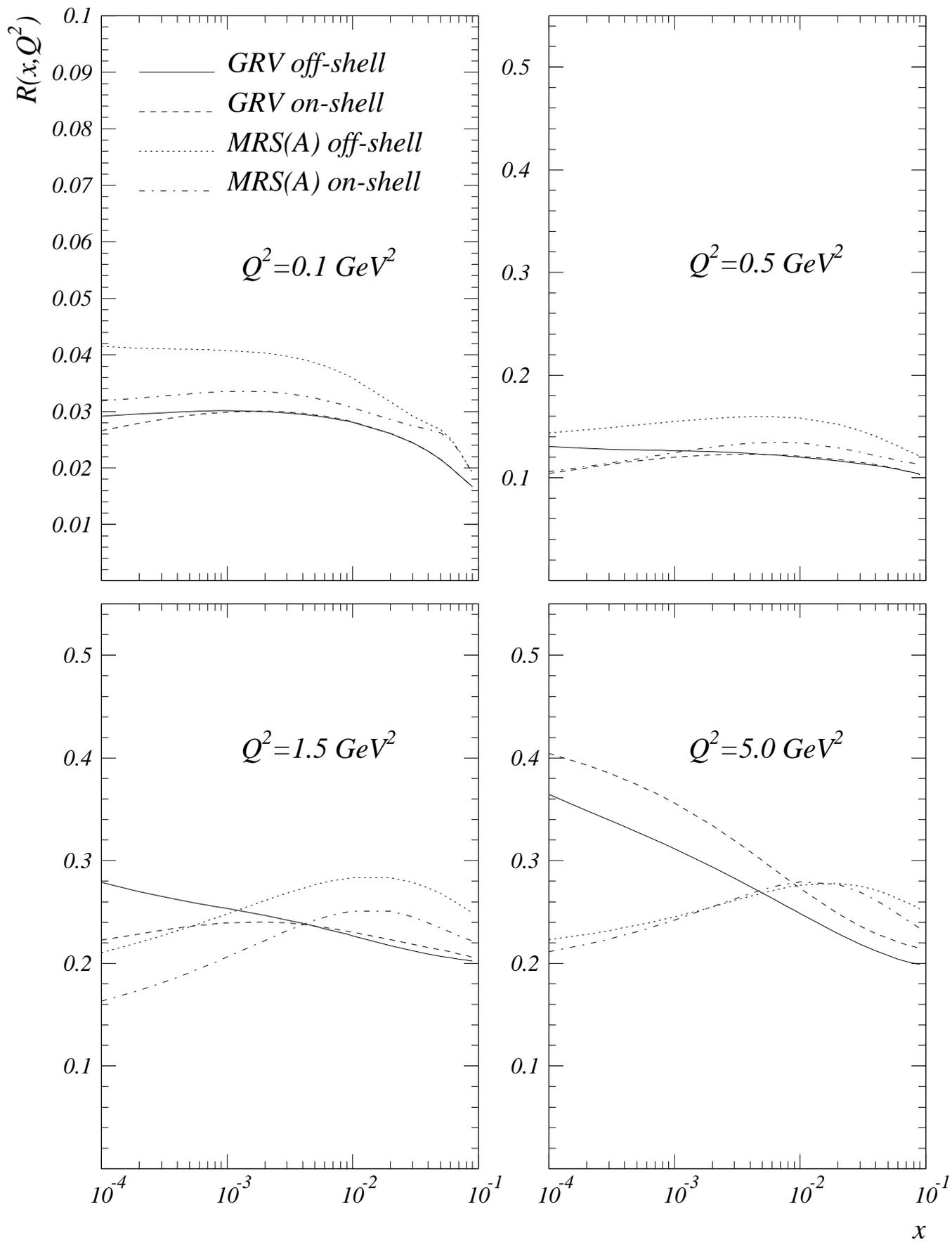
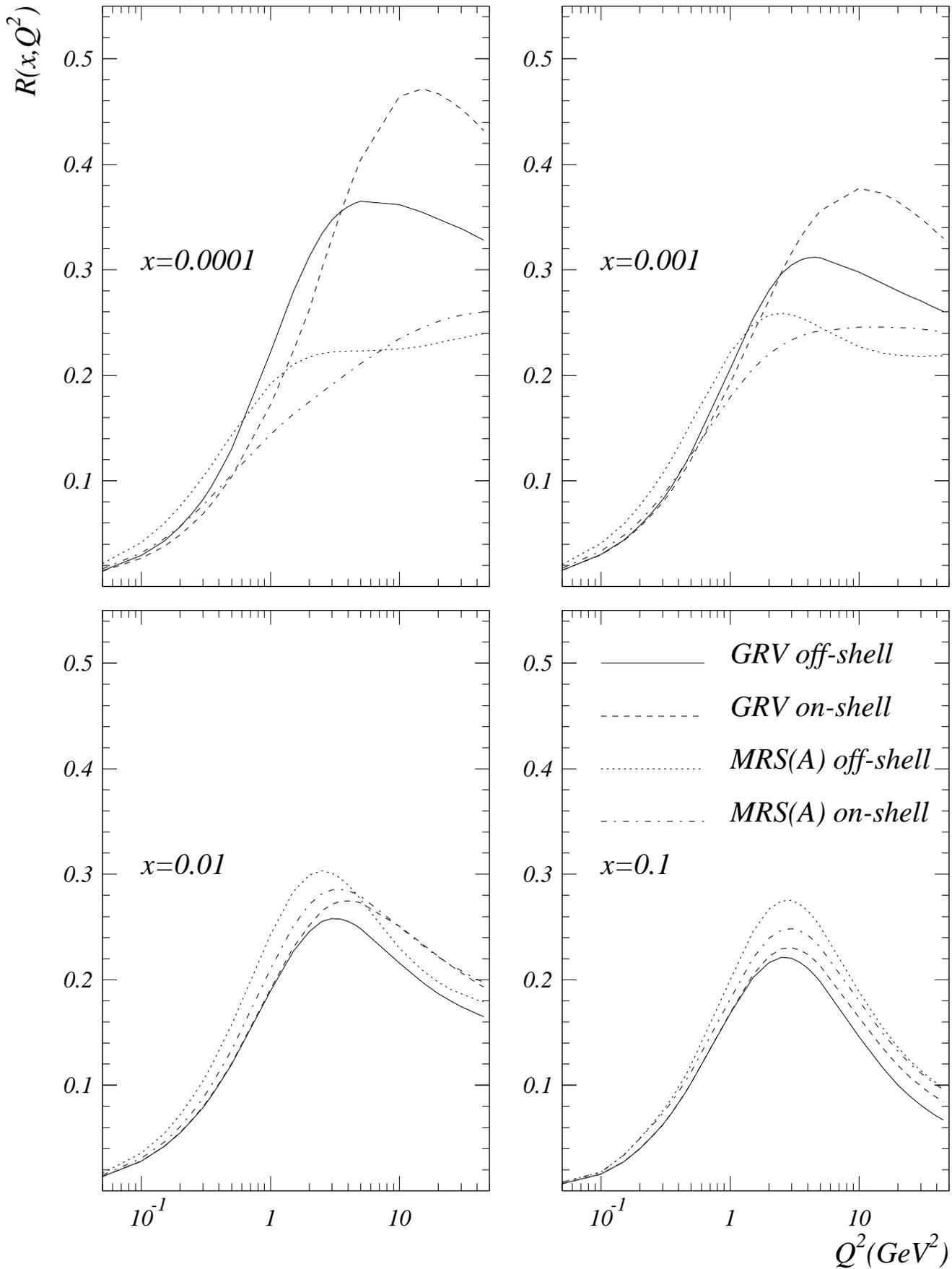
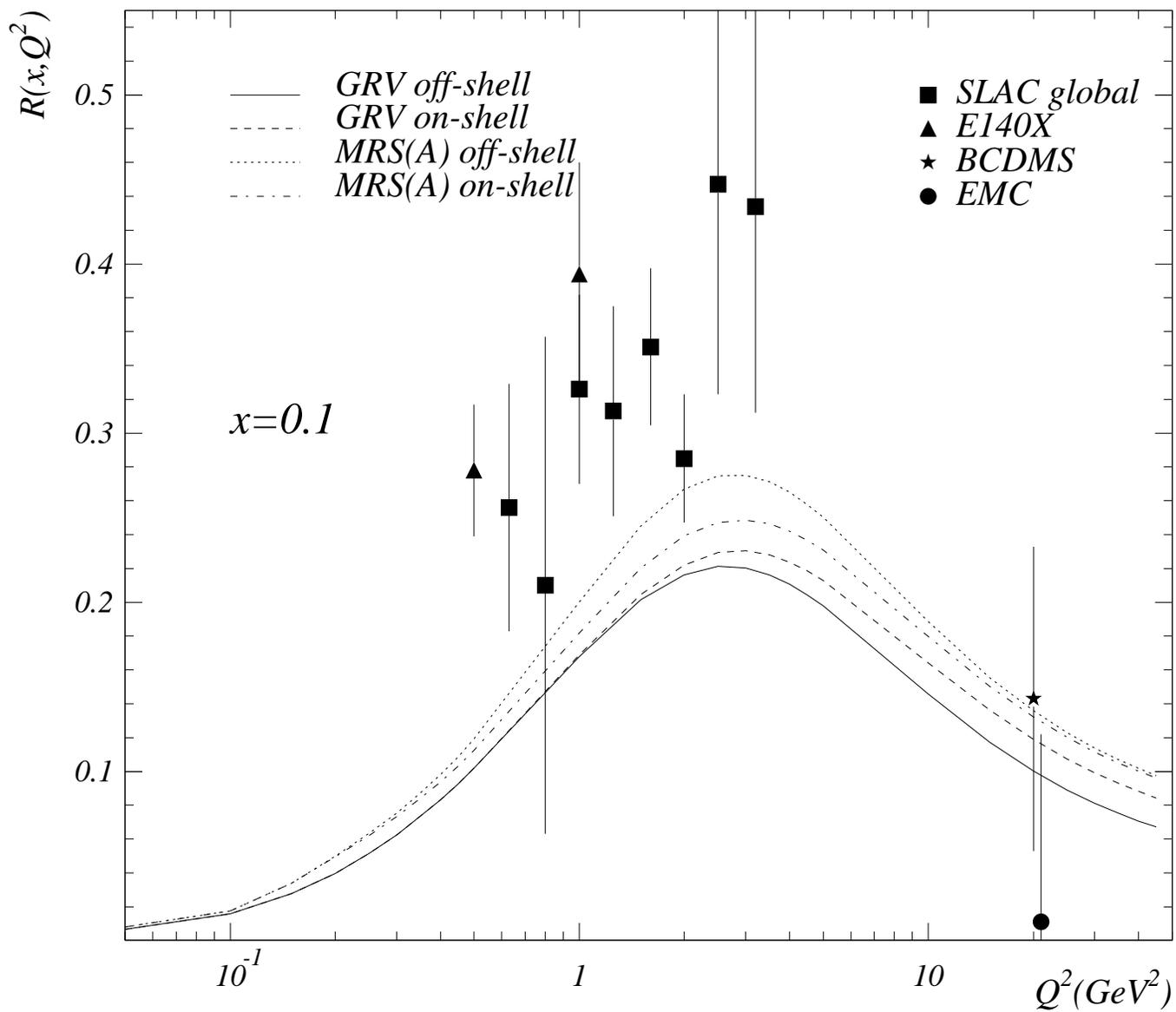
\end{document}